\documentclass[amssymb,prd,aps,amsmath,twocolumn,superscriptaddress,nofootinbib]{revtex4}
\usepackage{pslatex}
\usepackage{graphicx}
\usepackage{psfrag}

\usepackage[usenames]{color}
\usepackage{color,graphicx,epsfig,amsmath,amssymb}

\newcommand{\beq}{\begin{equation}}
\newcommand{\eeq}{\end{equation}}
\newcommand{\ben}{\begin{eqnarray}}
\newcommand{\een}{\end{eqnarray}}
\newcommand{\bi}{\begin{itemize}}
\newcommand{\ei}{\end{itemize}}

\newcommand{\ghost}[1]{ }

\begin{document}
\title{Can the morphology of $\gamma$-ray emission distinguish annihilating from decaying dark matter?}

\author{C\'eline B\oe hm}
\affiliation{LAPTH, Universit\'e de Savoie, CNRS, BP110, F-74941 Annecy-le-Vieux Cedex, France}
\email{celine.boehm@cern.ch}

\author{Timur Delahaye}
\affiliation{LAPTH, Universit\'e de Savoie, CNRS, BP110, F-74941 Annecy-le-Vieux Cedex, France}
\affiliation{Dipartimento di Fisica Teorica, 
  Universit\`a di Torino \& INFN - Sezione di Torino, 
  Via Giuria 1, I-10122 Torino, Italia.}
\email{delahaye@lapp.in2p3.fr}

\author{Joseph Silk}
\affiliation{Astrophysics department, Wilksinson Building,
Keble Road, OX1 3RH Oxford, UK}
\email{j.silk1@physics.ox.ac.uk}

\date{today}

\begin{abstract}
The recent results from the  PAMELA, ATIC, FERMI and  HESS experiments have focused attention on the possible existence of high energy cosmic ray  $e^+ e^-$ that may originate from dark matter (DM) annihilations or decays in the Milky Way.
 Here we examine the morphology of the $\gamma$-ray emission after propagation of the electrons generated by both  annihilating and decaying dark matter models. We focus on  photon energies of 1 GeV, 10 GeV, 50 GeV (relevant for the FERMI satellite) and consider different propagation parameters. Our main conclusion is that distinguishing annihilating from decaying dark matter may only be possible if the propagation parameters correspond to the most optimistic  diffusion models. In addition, we point to examples where morphology can lead to an erroneous  interpretation of the source injection energy. 
\end{abstract}
\maketitle

\section{Introduction}
Results from recent cosmic ray experiments (PAMELA \cite{pamela}, ATIC\cite{atic}, FERMI \cite{fermi}, HESS\cite{hess}) have raised the question of the origin of an ``anomalous'' population of high energy positrons in the Milky Way and motivated many studies. Correlation of the positron flux measured by  PAMELA with the $\gamma$-ray spectrum obtained by  FERMI  LAT is expected to give  insight into the injection energy of the high energy electron and positron ($e^+, e^-$) population, and should also probe their spatial and energy distribution. 
In scenarios where high energy $e^+, e^-$ are emitted by dark matter (DM), the spatial and energy distribution of this ``additional'' cosmic ray population is expected to follow the DM halo distribution at the injection energy $E=E_{inj}$. This implies (assuming a spherical DM halo) that they should be spherically distributed with an energy density that is maximal near the Galactic Centre. 

However, this picture could be modified if the high energy $e^+, e^-$ spatially propagate and lose energy in the galaxy owing to inverse Compton and synchrotron losses. As a consequence of propagation, not only will the spatial and energy distributions of the high energy $e^+, e^-$ be modified but  their final energy will be smaller than $E_{inj}$.  The $\gamma$-ray spectrum  obtained after propagation could therefore differ significantly from that obtained at injection. 

The issue of the $\gamma$-ray flux associated with DM annihilations or decays into leptons has been addressed in several papers. For example, both the $\gamma$-ray flux and $\gamma$-ray spectrum in decaying and annihilating scenarios have been discussed in ref.~\cite{cirelli1,cirelli2} but propagation was actually neglected. More recently, the authors of ref.\cite{sigl} have predicted the expected $\gamma$-ray flux in a decaying DM model, taking into account $e^+, e^-$ propagation. Although computation of the flux is important, exploiting its value will be difficult owing to large uncertainties due to  astrophysical sources at these energies (Ref.\cite{stefano}). Other papers have considered specific positions on the sky (e.g. intermediate galactic latitudes, \cite{bertoneGC,ullio2}), or rely on very large-scale anisotropies (\cite{sigl}). The work in ref. \cite{ullio1} raised the question of the morphology of the $\gamma$-ray emission but mainly focused on the spectrum; however the propagation parameters adopted are not those favoured by  MCMC studies  (\cite{antje}). 

The question we raise in this Letter is  whether the morphology  of the $\gamma$-ray emission alone (rather than the flux) can actually help to discriminate between the different DM scenarios.
To address this issue, we compute $\gamma$-ray maps originating 
from the interactions of $e^+ e^-$ with the Interstellar Radiation Field (ISRF) spectra after propagation. We assume that the dark matter only annihilates or decays  into   $e^+ e^-$ pairs and focus on $\gamma$-ray energies that are accessible by the FERMI satellite, namely $E_{\gamma}= 1,10,50$ GeV. {We have neglected prompt $\gamma$ emission which could arise from internal bremsstrahlung because its spectrum is model-dependent, however one should keep in mind that for some models, this emission could modify our conclusions.}
Given our assumptions, the injection energy of the $e^+$ and $e^-$  corresponds to either the DM mass $m_{dm}$ or half the DM mass ($E_{inj} = m_{dm}$ or $m_{dm}/2$), depending on whether DM is annihilating or decaying respectively. We will consider three values of the injection energy: $E_{inj} = 100, 500, 1000$ GeV. At given $E_{inj}$, the comparison between decaying and annihilating scenarios is immediate. In addition, since there are quite large uncertainties in the propagation parameters of cosmic rays, we will use three different sets of parameters referred to as (MIN,MED,MAX) (cf Ref.~\cite{2001ApJ...555..585M,2001ApJ...563..172D}), which give a fair idea of the related uncertainty. We present difference maps of the $\gamma$-ray contributions that highlight how morphology could help discriminate between the competing models.


\section{Producing gamma-ray maps \label{map_obt}}
 
To generate these maps, we apply the propagation scheme introduced by \cite{2001ApJ...555..585M}, modified according to \cite{2008PhRvD..77f3527D,2006ApJ...648L..29P} for primary electrons. We compute the halo function $\tilde{I}$ in terms of the electron energy:
\begin{eqnarray}
\tilde{I} = \sum_i \sum_n \ J_0\left(\frac{\alpha_i r}{R_{\rm{gal}}}\right) \times \varphi_n(z) \times e^{\left\{-\left(\frac{n \pi}{2L}\right)^2 \ + \ \frac{\alpha_i^2}{R_{\rm{gal}^2}} \right\} \lambda_d^2} \times Q_{in} \ \label{Itild}
\end{eqnarray} 
where $Q_{in} = \kappa \ R_{in}$ is the Fourrier-Bessel transform of the source term (with $\kappa = \frac{{\cal{A}}}{\eta} \times (\rho_0/m_{\rm{dm}})^\zeta $ and ${\cal{A}}=\sigma v$ the annihilation cross section if $\zeta=2$ and ${\cal{A}}= \Gamma$ the decay rate if $\zeta=1$; $\eta$ is the multiplicity, i.e. $=2$ if Majorana particles, $1$ otherwise), 
$L$ is diffusion slab half-thickness,
$R_{gal}$ is Galaxy radius and $\lambda_d$ the propagation length expressed as:
$$\lambda_d = 4 \times K_0 \times \int_{E_{min}}^{E_{inj}} \frac{E^{\delta}}{b(E)} dE$$
where $b(E)$ is the loss term and $K_0, \delta$ the diffusion parameters. Note that in  Eq.~(\ref{Itild}), $r,z$ are cylindrical coordinates and $\alpha_i$ are the roots of the Bessel function $J_0$. The gamma ray flux detected at the Earth is given by the integration along the line of sight of the convolution of the electron flux with the gamma emissivity of the electron interacting with the Interstellar Radiation Field (ISRF). The ISRF is mainly made of stellar light which is absorbed and re-emitted in the infrared by galactic dust. 
We use the model from \cite{2008ApJ...682..400P} which can be fitted by a sum of blackbody (BB)-like spectra as in \cite{TIJU}. Although this fit is valid only in the two kpc around the Sun, we expect that only the relative amplitudes associated with each BB vary from one position to another, but the temperatures should remain the same.
The emissivity is computed in the same way as the losses in \cite{TIJU}, making the approximation that the outgoing photon spectrum is a delta function (\cite{BG}) for each blackbody with which the electrons are interacting. 

\section{results \label{results}}

\begin{table}

\begin{center}

\begin{tabular}{|c|c|c|c|c|c|c|c|}
\hline
$E_{inj} = 100$GeV & parameter & $a_{10}$ & $b_{10}$ & $\epsilon_{10}$  &$a_{2}$ & $b_{2}$ & $\epsilon_{2}$\\
$E_{\gamma}=10$ GeV & & & & & & &\\
\hline
\hline
\hline
ann  &MIN & 19 & 6.5 & 0.66 &7.5 &4 &0.47\\
decay & MIN & 95 & 8 & 0.92 &19 &4 &0.79\\
\hline
 ann &MED & 10 & 25 & 0.31 &15 &8.5 &0.1\\
 decay &MED & 126 & 35 & 0.72 &30.5 &15.5 &0.49\\
\hline
ann  &MAX & 54 & 45.5 & 0.16 &23 &21.5 &0.07\\
decay & MAX & 179.5 & 67 & 0.63 &40.5 &32.5 &0.2\\
\hline
\end{tabular}

\caption{Ellipticity for $E_{\gamma}=$ 10 GeV and $E_{inj} = 100$ GeV for 0.1 ( subscript $10$) and 0.5  ( subscript $2$) of the intensity.}

\label{tab:particles}\end{center}

\end{table}

\begin{figure*}
	\centering
		\includegraphics[width=16.5cm]{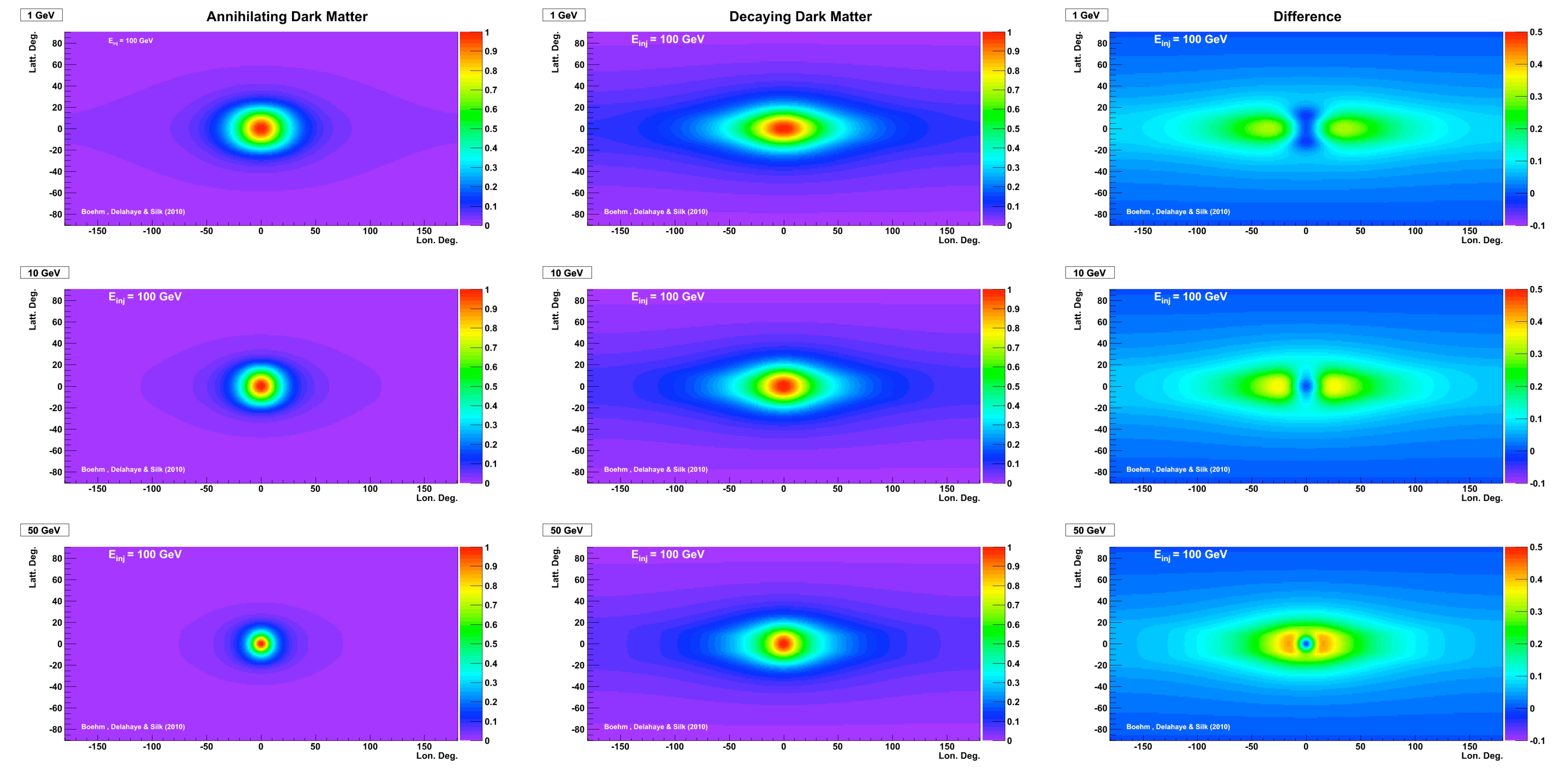}
	\caption{Annihilating versus decaying DM for $E_{inj}=$100 GeV and $E_{\gamma}=1,10,50$ GeV. In these figures, fluxes are normalized to the central bin so as to make the comparison of propagation length obvious. Ellipticities at 0.1 of the central bin intensity are equal to $\epsilon_{10}=0.41, 0.31,0.15$ and $\epsilon_{10}=0.75,0.72,0.68$ for annihilating versus decaying DM respectively.}
	\label{fig:Fig1}
\end{figure*}

We now present the pixelized maps (with a pixel size of 1 square degree) that we have obtained for the different scenarios. Fig.~\ref{fig:Fig1} illustrates (for both annihilating and decaying dark matter models) the difference between the propagation patterns that arise by fixing $E_{inj}$ to 100 GeV and considering three gamma ray energies $E_{\gamma}= 1, 10, 50$ GeV.  As one can see, in both cases, the $e^+, e^-$ which 
give rise to 1 GeV photons have propagated further than those giving rise to 50 GeV photons. These features are common to all the maps including those obtained for heavier dark matter candidates.
Note that the propagation parameters that we have considered to obtain this map correspond to the MED set ($L=4$ kpc, $\delta=0.7$, $K_0= 0.0112$ $\rm{kpc}^2/\rm{Myr}$).

\begin{figure*}
	\centering
		\includegraphics[width=16.5cm]{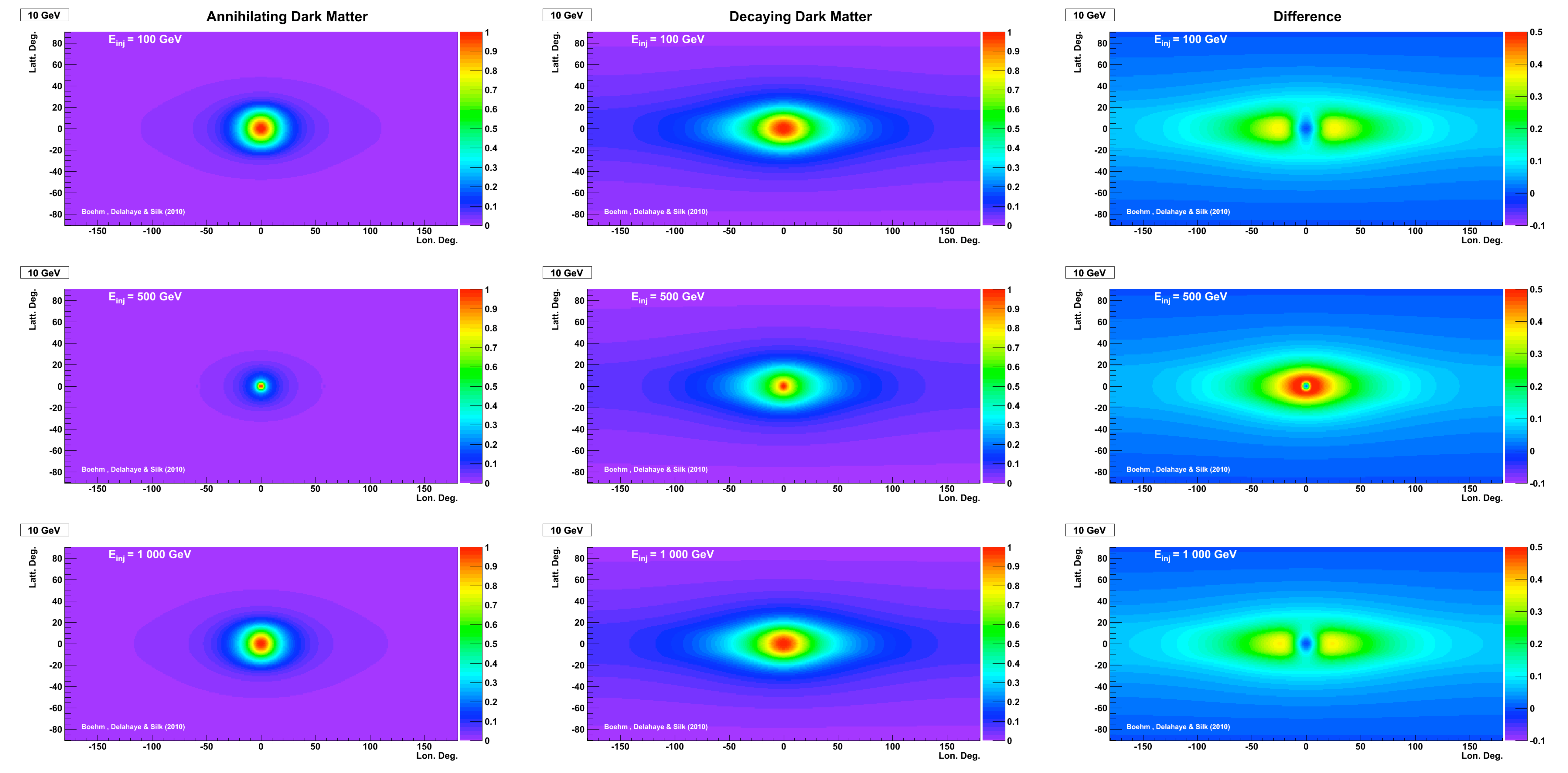}
	\caption{Annihilating versus decaying DM for $E_{\gamma}=10$ GeV and $E_{inj}=100,500,1000$ GeV. Ellipticities at 0.1 of the central bin intensity  are equal to $\epsilon_{10}=0.31, 0.06,0.35$ and $\epsilon_{10}=0.72,0.67,0.72$ for annihilating versus  decaying DM respectively.}
	\label{fig:Fig2}
\end{figure*}

In Fig.\ref{fig:Fig2}, we have fixed $E_{\gamma}$ to 10 GeV and considered  three values of $E_{inj}$ (for both annihilating and decaying DM). Interpreting the features for the particular case $E_{inj} = 500 GeV$ and $E_{\gamma} = 10 GeV$ is non-trivial. As can be seen from Table~\ref{tab:particles}, because  the ISRF is made of more than one BB, $\gamma$--emission at 10~GeV can actually be due to more than one electron population. Indeed  bright emission at 10~GeV could be due either to electrons of $\sim$20~GeV interacting with UV light or to $\sim$500~GeV electrons interacting with IR light. Hence as seen 
 in Fig.\ref{fig:Fig2}, the 10~GeV emission is nearly spherical, and could be interpreted either as an injection energy of $\sim$20~GeV or of $\sim$500~GeV,  leading to very different interpretations concerning the mass of the DM particle. However this degeneracy can be lifted by looking at higher energies, as  electrons injected at 20~GeV cannot produce gamma rays of 50~GeV.
This threshold effect stresses how important it is to look at different $\gamma$-ray energies and to compare the various morphologies in order  to understand the properties of the DM.

In the third column, we exhibit the difference between the two normalized maps (decays $-$ annihilations) so as to exhibit the differences of morphology between these two emission models. The negative values at the Galactic Centre confirm that the $e^+, e^-$ from annihilating DM are mainly produced locally and that propagation cannot  completely smooth out the contrast with respect to decaying DM electrons and positrons.

\begin{table}[h!]

\begin{center}

\begin{tabular}{|c|c|c|c|}
\hline
BB & E$_\gamma$ = 1 GeV & E$_\gamma$ = 10 GeV & E$_\gamma$ = 50 GeV\\
\hline
\hline
CMB & 527 GeV & 1.7 TeV & 3.7 TeV\\
\hline
IR & 151 GeV & 479 GeV & 1.1 TeV\\
\hline
Stellar & 49 GeV & 155 GeV & 348 GeV\\
\hline
UV1 & 15 GeV & 48 GeV & 107 GeV\\
\hline
UV2 & 11 GeV & 35 GeV & 78 GeV\\
\hline
UV3 & 6 GeV & 18 GeV & 40 GeV\\
\hline
\end{tabular}

\caption{Electron energy E$_e$ responsible for the emission of an electron of energy E$_\gamma$ through  inverse Compton scattering on each blackbody component of the ISRF.}

\label{tab:particles}\end{center}

\end{table}

\begin{figure*}
	\centering
		\includegraphics[width=16.5cm]{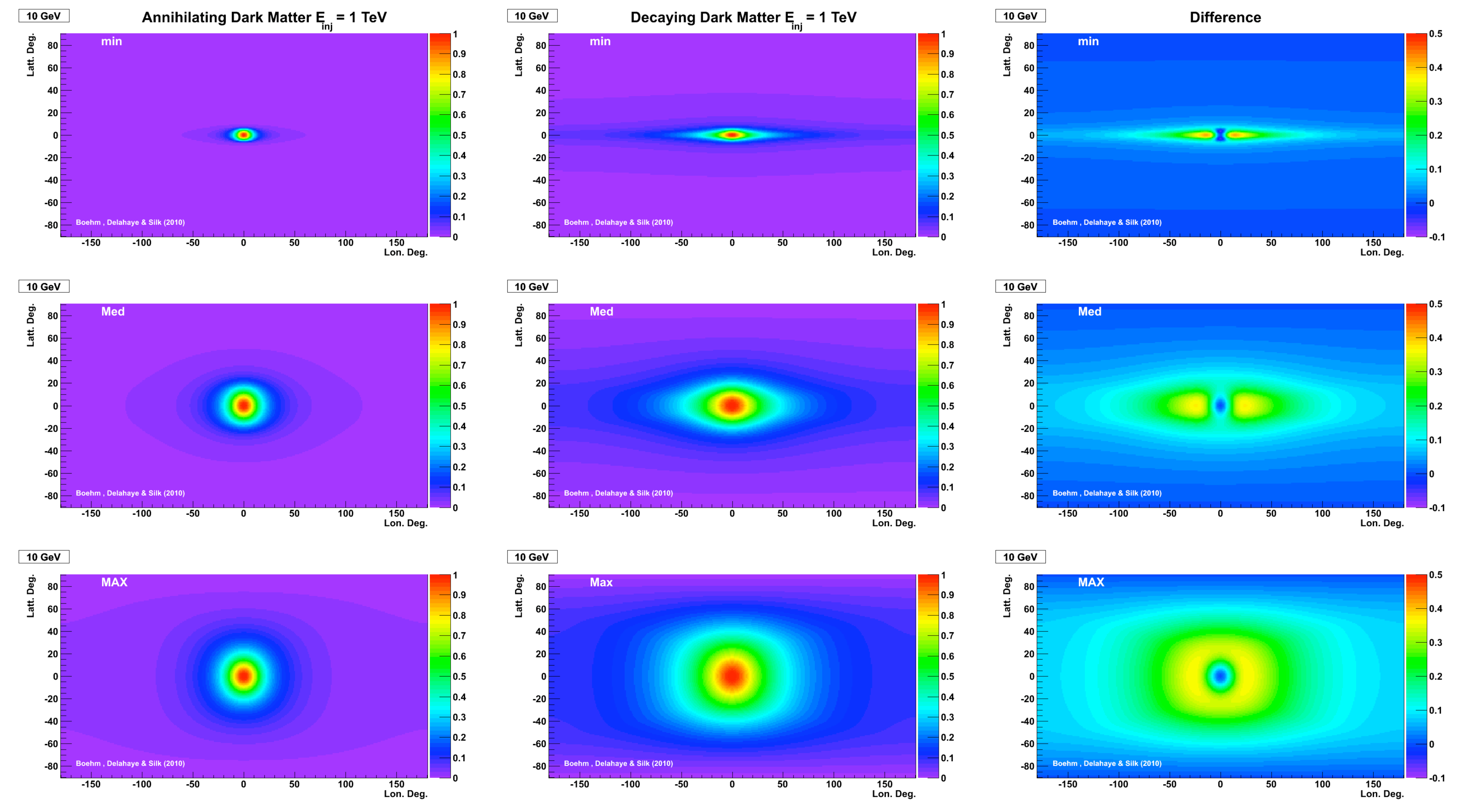}
	\caption{Annihilating versus decaying DM for $E_{inj}=$1 TeV and the min and MAX propagation parameters. Ellipticities at 0.1 of the central bin intensity for MIN and MAX are equal to $\epsilon_{10}=0.67, 0.35,0.15$ and $\epsilon_{10}=0.92,0.72,0.58$ for annihilating versus  decaying DM respectively.}
	\label{fig:Fig3b}
\end{figure*}

  In Fig.\ref{fig:Fig3b}, we show the effect of the propagation parameters for $E_{inj}=$~1~TeV. As is expected, the $e^+ e^-$ diffuse far more for the set of propagation parameters MAX (for which $L = 15$ kpc) than for MIN. Although it may be possible to constrain decaying versus annihilating DM in the MAX and MED cases, it seems impossible to distinguish these two scenarios in the MIN case.
To compare the propagation features between annihilating and decaying scenarios, 
it is useful to look at the ellipticities $\epsilon_{10}$ and $\epsilon_{2}$ of the $\gamma$--emission. To define these quantities, we measure the size of the semi-major axis $a_{10}$ (or $a_{2}$) and the semi-minor axis $b_{10}$ (or $b_{2}$) of the ellipse that has an intensity of one tenth (or one half) of the maximal intensity. Ellipticity is then defined as $1-b/a.$
 The results are summarized in the caption of Fig.~\ref{fig:Fig1},\ref{fig:Fig2},\ref{fig:Fig3b} and Table~\ref{tab:particles}. Only
for larger masses, or in the optimistic case where the sensitivity allows us to measure $\epsilon_{10}$, is  discrimination
possible, especially for the 
 MAX propagation model.

\section{Conclusion \label{conclusion}}
We have generated maps of $\gamma$-ray emission associated with $e^+, e^-$ population originating from DM annihilations or decays. We show that propagation is important for both  DM scenarios, but although the propagation features differ, they are  difficult to distinguish if the propagation parameters correspond to MIN (and perhaps MED) rather than to MAX. This is, in fact,  surprising, as one might have expected these two scenarios, which involve distinct powers of the dark matter density,  to differ significantly. 
Actually, in the MIN case, detection would  be extremely challenging since most of  the signal would be hidden by galactic sources. In some cases,  the IRSF can make the Galactic Centre  bright enough to  be misinterpreted as $e^+ e^-$ with   a lower injection energy.\\
We have verified that changing the energy density of the ISRF has little effect as the increase of the $\gamma$-ray  emissivity is partially compensated by the electron density decrease due to increased energy losses. Varying the intensity of the magnetic field within reasonable values has also little impact as synchrotron emission is not the main energy loss term in most  cases. In both cases the impact is mainly on the intensity and not on the ellipticity. However a full spatial description of both the ISRF and the magnetic field could have effects that are beyond the scope of  our analytical approach. 

\vspace{0.2cm}

We would like to thank P. Salati and A. Fiasson for very useful discussions.
 and acknowledge fundings from the "low energy electron and positron propagation" PICS.

\bibliography{timur}
\end{document}